\documentstyle[11pt]{article}
\input epsf
\begin{document}
\title{\large\bf Properties of
the Scalar Mesons $f_0(1370)$, $f_0(1500)$ and $f_0(1710)$}

\author{\small De-Min Li${}^b$\footnote{ E-mail:
lidm@alpha02.ihep.ac.cn},~ Hong Yu${}^{a,b,c}$, Qi-Xing Shen${}^{a,b,c}$\\
\small \em $^a$ CCAST (World Lab), P.O.Box 8730, Beijing 100080, P.R. China\\
\small \em $^b$ Institute of High Energy Physics, Chinese Academy of Sciences,\\
\small \em P.O. Box 918 (4), Beijing 100039, P.R. China\footnote{Mailing address}\\
\small \em $^c$ Institute of Theoretical Physics, Chinese Academy of Sciences, Beijing 100080, P.R. China}
\date{}
\maketitle
\vspace{0.5cm}

\begin{abstract}
In the three-state mixing framework, considering the possible
glueball components of $\eta$ and $\eta^\prime$, we
investigate the hadronic decays of
$f_0(1370)$, $f_0(1500)$ and $f_0(1710)$ into two pseudoscalar mesons. The
quarkonia-glueball content of the three states is determined from the fit
to the new data presented by the WA102 Collaboration. We find that these data are
insensitive to the possible glueball components of $\eta$ and
$\eta^\prime$. Furthermore, we discuss some
properties of the mass matrix describing the mixing of the isoscalar scalar
mesons.
\end{abstract}

\vspace{0.5cm}



\newpage
\baselineskip 24pt
\section{Introduction}
\indent

Recently, based on the mass matrix motivated by Ref.\cite{Wein},
Ref.\cite{4241} has
investigated the implications of the new data presented by the WA102
Collaboration\cite{102} for the glueball-quarkonia content of
$f_0(1370)$, $f_0(1500)$ and $f_0(1710)$ ($f_1$, $f_2$ and $f_3$
respectively stand for
$f_0(1370)$, $f_0(1500)$ and $f_0(1710)$ below).  We
propose that some points can be improved on. First, in
the reduced
partial width $\Gamma(f_i\rightarrow\eta\eta^\prime)$ in Ref.\cite{4241},
the sign of the contribution of the diagram iv) of Fig. 1 was flipped, it
should be negative. The flipped sign actually
arose from a typo in the equation (A5) of Ref.\cite{A5} where the
$\lambda$
should all read $1/\lambda$\footnote{We wish to thank F.E.
Close
and A. Kirk for useful discussions on this matter.}. Second, the
mixing angle
of $\eta$ and $\eta^\prime$ was determined to be a very small value of
$-5\pm4^\circ$\cite{4241} which is inconsistent with the value of
$-15.5\pm1.5^\circ$ determined from a rather exhaustive and up-to-date
analysis of data including strong decays of tensor and higher spin mesons,
electromagnetic decays of vector and pseudoscalar mesons, and the decays
of $J/\psi$\cite{Bramon}.
Also, the possibility that the glueball components exist in $\eta$ and
$\eta^\prime$ was not considered in Ref.\cite{4241}.
Refs.\cite{pseud} already suggested that the $\eta$
and $\eta^\prime$ wave functions need glueball components.

In this work, instead of the mass matrix in which one should
confront with the confused mass level order about the masses of the
bare states $(u\overline{u}+d\overline{d})/\sqrt{2}$, $s\overline{s}$
and glueball\cite{Wein,order,CLOSE}, we shall adopt another
mixing scheme which can be
related to the mass matrix to describe the mixing of $f_1$, $f_2$ and
$f_3$, then we can discuss some properties of the mass matrix based on
our preferred results. In addition, we shall consider the possibility that
the glueball components exist in $\eta$ and $\eta^\prime$ when we
investigate the hadronic decays of $f_1$, $f_2$ and $f_3$ into two
pseudoscalar mesons, and check whether these new data are
sensitive to the possible glueball components of $\eta$ and $\eta^\prime$
or not.

\section{Mixing scheme and decays}
\indent

Based on three Euler angles
$\theta_1$, $\theta_2$ and $\theta_3$, the mixing of $f_1$, $f_2$ and
$f_3$ can be described as:
\begin{equation}
\left(\begin{array}{c}
f_1\\
f_2\\
f_3
\end{array}\right)
=\left(\begin{array}{ccc}
a_8&a_1&a_g\\
b_8&b_1&b_g\\
c_8&c_1&c_g
\end{array}\right)
\left(\begin{array}{c}
|8\rangle\\
|1\rangle\\
|G\rangle
\end{array}\right)=
\left(\begin{array}{ccc}
x_1&y_1&z_1\\
x_2&y_2&z_2\\
x_3&y_3&z_3
\end{array}\right)
\left(\begin{array}{c}
|N\rangle\\
|S\rangle\\
|G\rangle
\end{array}\right)
\end{equation}
with
\begin{eqnarray}
&\left(\begin{array}{ccc}
a_8&a_1&a_g\\
b_8&b_1&b_g\\
c_8&c_1&c_g
\end{array}\right)=
\left(\begin{array}{ccc}
c_1c_2c_3-s_1s_3&-c_1c_2s_3-s_1c_3&c_1s_2\\
s_1c_2c_3+c_1s_3&-s_1c_2s_3+c_1c_3&s_1s_2\\
-s_2c_3&s_2s_3&c_2
\end{array}\right),\\
&\left(\begin{array}{ccc}
x_1&y_1&z_1\\
x_2&y_2&z_2\\
x_2&y_3&z_3
\end{array}\right)=
\left(\begin{array}{ccc}
a_8&a_1&a_g\\
b_8&b_1&b_g\\
c_8&c_1&c_g
\end{array}\right)
\left(\begin{array}{ccc}
\sqrt{\frac{1}{3}}&-\sqrt{\frac{2}{3}}&0\\
\sqrt{\frac{2}{3}}&\sqrt{\frac{1}{3}}&0\\
 0&0&1
\end{array}\right),
\end{eqnarray}
where
$|8\rangle=|u\overline{u}+d\overline{d}-2s\overline{s}\rangle/\sqrt{6}$,
$|1\rangle=|u\overline{u}+d\overline{d}+s\overline{s}\rangle/\sqrt{3}$,
$|N\rangle=|u\overline{u}+d\overline{d}\rangle/\sqrt{2}$,
$|S\rangle=|s\overline{s}\rangle$, $|G\rangle=|gg\rangle$;
$c_1$ ($c_2,~c_3$)$\equiv\cos\theta_1$
($\cos\theta_2,~\cos\theta_3$),
$s_1$ ($s_2,~s_3$)$\equiv\sin\theta_1$
($\sin\theta_2,~\sin\theta_3$), and $-180^\circ\leq\theta_1\leq
180^\circ$,~$0^\circ\leq\theta_2\leq
180^\circ$,~$-180^\circ\leq\theta_3\leq 180^\circ$.
One advantage of this mixing model is only 3 unknown parameters with the
definite changing ranges.

Considering that the glueball components possibly exist in the final
isoscalar pseudoscalar mesons\cite{pseud},
for the hadronic decays of $f_i$ (here and below, $i=1,~2,~3$) into
pseudoscalar meson
pairs, we consider the following coupling modes as indicated in Fig. 1: i)
the coupling of the $q\overline{q}$ components of $f_i$ to the
final pseudoscalar meson pairs, ii) the coupling of the glueball
components of $f_i$ to the final pseudoscalar meson pairs via
$qq\bar{q}\bar{q}$ intermediate states, iii) the coupling of the glueball components of $f_i$ to the
glueball components of the final isoscalar pseudoscalar mesons, and iv) the
coupling
of the glueball components of $f_i$ to the $q\overline{q}$ components of
the final isoscalar pseudoscalar meson pairs.  Based on these coupling
modes, the effective Hamiltonian describing the hadronic decays of $f_i$
into two pseudoscalar mesons can be written as\cite{Hamil}
\begin{eqnarray}
&&H_{eff}=g_1{\bf Tr}(f_{F}P_FP_F)+g_2f_{G}{\bf
Tr}(P_FP_F)+g_3f_{G}P_GP_G+g_4f_{G}{\bf Tr}(P_F){\bf Tr}(P_F),
\end{eqnarray}
where $g_1$, $g_2$, $g_3$ and $g_4$ describe the effective
coupling strengths of the coupling modes i), ii), iii) and iv),
respectively. $f_{G}$ and $P_G$ are $SU(3)$ flavor singlets describing the
glueball components of $f_i$ and the final isoscalar pseudoscalar mesons,
respectively. $f_{G}$ and $P_G$ can be given by \begin{eqnarray}
f_{G}=\sum\limits_{i}z_if_i,~P_G=\sum\limits_{j}z_jj,
\end{eqnarray}
where $z_j$ denotes the glueball content of $j$ (here and below
$j=\eta$, $\eta^\prime$). $f_{F}$ and $P_F$ are $3\times3$ flavor matrixes
describing the
$q\bar{q}$ components of $f_i$ and the final pseudoscalar mesons,
respectively. $f_{F}$ can be written as
\begin{equation}
f_{F}=\left(\begin{array}{ccc}
\sum\limits_{i}\frac{x_i}{\sqrt{2}}f_i&0&0\\
0&\sum\limits_{i}\frac{x_i}{\sqrt{2}}f_i&0\\
0&0&\sum\limits_{i}y_if_i
\end{array}\right),
\end{equation}
$P_F$ can be written as
\begin{equation}
P_F=\left(\begin{array}{ccc}
\frac{1}{\sqrt{2}}\pi^0+\sum\limits_{j}\frac{x_j}{\sqrt{2}}j&\pi^{+} &K^{+}\\
\pi^{-}&-\frac{1}{\sqrt{2}}\pi^0+\sum\limits_{j}\frac{x_j}{\sqrt{2}}j
&K^0\\
K^{-}&\overline{K^0}&\sum\limits_{j}y_jj
\end{array}\right),
\end{equation}
where $x_j$ and $y_j$ denote the
$(u\bar{u}+d\bar{d})/\sqrt{2}$, $s\bar{s}$ contents of $j$,
respectively, and they
satisfy $x^2_j+y^2_j+z^2_j=1.$

Introducing $g_2/g_1=r_1$, $g_3/g_1=r_2$, $g_4/g_1=r_3$, from Eqs.
(4)$\sim$(7), one can obtain
\begin{eqnarray}
&&\Gamma{(f_i\rightarrow\pi\pi)}=
3g^2_1q_{i\pi\pi}[x_i+\sqrt{2}z_ir_1]^2,
\\
&&\Gamma{(f_i\rightarrow K\bar{K})}=
g^2_1q_{iK\bar{K}}[x_i+\sqrt{2}y_i+2\sqrt{2}z_ir_1]^2,
\\
&&\Gamma{(f_i\rightarrow\eta\eta)}=
g^2_1q_{i\eta\eta}[x^2_\eta x_i+\sqrt{2}y^2_\eta y_i
+\sqrt{2}(x^2_\eta+y^2_\eta)z_ir_1
\nonumber\\
&&~~~~~~~~~~~~~~~~~~+\sqrt{2}z^2_\eta
z_ir_2+(2\sqrt{2}x^2_\eta+\sqrt{2}y^2_\eta+4x_\eta
y_\eta)z_ir_3]^2,
\\
&&\Gamma{(f_i\rightarrow\eta\eta^\prime)}=
g^2_1q_{i\eta\eta^\prime}[\sqrt{2}x_\eta x_{\eta^\prime}x_i+2y_\eta
y_{\eta^\prime}y_i
+2(x_\eta x_{\eta^\prime}+y_\eta
y_{\eta^\prime})z_ir_1
\nonumber\\
&&~~~~~~~~~~~~~~~~~~+2z_\eta z_{\eta^\prime}z_ir_2+2(2x_\eta
x_{\eta^\prime}+\sqrt{2}x_\eta
y_{\eta^\prime}+\sqrt{2}x_{\eta^\prime}y_\eta+y_\eta
y_{\eta^\prime})z_ir_3]^2,
\end{eqnarray}
where $q_{iP_1P_2}$ is the decay momentum for the decay mode
$f_i\rightarrow P_1P_2$,
\begin{eqnarray}
q_{iP_1P_2}=\sqrt{[M^2_i-(M_{P_1}+M_{P_2})^2][M^2_i-(M_{P_1}-M_{P_2})^2]}/2M_i,
\end{eqnarray}
$M_i$ is the mass of $f_i$, $M_{P_1}$ and $M_{P_2}$ are the masses of the final pseudoscalar mesons
$P_1$ and $P_2$, respectively, and we take $M_K=\sqrt{(M^2_{K^{\pm}}+M^2_{K^0})/2}$.

For $\Gamma(f_i\rightarrow\eta\eta)$ and
$\Gamma(f_i\rightarrow\eta\eta^\prime)$, the contribution of the coupling
mode iv) given in our present work differs from that given in
Ref.\cite{4241} since we don't adopt the assumption employed by
Ref.\cite{4241} that the coupling
of the glueball components of $f_i$ to the $q\bar{q}$ components of the
isoscalar pseudoscalar mesons occurs dominantly through their $s\bar{s}$
content in chiral symmetry. In addition, even under this assumption (i.e.,
$x_j$ in the terms containing $r_3$ is set to be zero),
for $\Gamma(f_i\rightarrow\eta\eta^\prime)$, the contribution of
the mode iv) should be proportional to $+y_{\eta}y_{\eta^\prime}$ but not
$+2\alpha\beta\equiv -y_{\eta}y_{\eta^\prime}$ given by Ref.\cite{4241}.

\section{Fit results}
\indent

Before performing the fit to determine the glueball-quarkonia content of
$f_i$, we should first determine the parameters $x_j$, $y_j$
and $z_j$. We
will adopt the mixing scheme mentioned above to discuss the mixing of
$\eta$, $\eta^\prime$ and $\eta(1410)$. Recently,
the mixing of the three states based on a mass matrix has been discussed
in Ref.\cite{DML}. Based on the equations (22)$\sim$(29) in Appendix A,
the $\theta_1$, $\theta_2$ and $\theta_3$ are determined as
$\theta_1=-98^\circ$, $\theta_2=30^\circ$ and $\theta_3=-95^\circ$,
and $x_j$, $y_j$ and $z_j$ are determined as
\begin{eqnarray}
&&x_\eta=-0.731,~y_\eta=0.679,~z_\eta=-0.069,~x_{\eta^\prime}=-0.566,
~y_{\eta^\prime}=-0.660,~z_{\eta^\prime}=-0.495,
\end{eqnarray}
with $\chi^2=1.64$, which is consistent with the results given by
Refs.\cite{DML}.
If we set $\theta_2$ and $\theta_3$ to be zero, i.e., we do not consider
the possible glueball components of $j$, the mixing angle of $\eta$ and
$\eta^\prime$ is determined
to be the value of $-15^\circ$ which is in good agreement with
the value of $-15.5\pm1.5^\circ$ given by\cite{Bramon}, and $x_j$
and $y_j$ are determined as
\begin{eqnarray}
&&x_\eta=y_{\eta^\prime}=[\cos(-15^\circ)-\sqrt{2}\sin(-15^\circ)]/\sqrt{3},\\
&&x_{\eta^\prime}=-y_\eta=[\sin(-15^\circ)+\sqrt{2}\cos(-15^\circ)]/\sqrt{3},
\end{eqnarray}
with $\chi^2=9.19$. The $\chi^2$ implies that the
$\eta$ and $\eta^\prime$ wave functions need the additional glueball
components. The predicted and measured results are shown in Table I.

In order to investigate whether the new data given by Ref.\cite{102} are
sensitive to the possible glueball components of
$\eta$ and $\eta^\prime$ or not, we perform the fit to the data
presented in Table III in two cases: a)
$z_j\not=0$ and b) $z_j=0$.
In the fit procedure , we take $M_1=1.312$ GeV, $M_2=1.502$ GeV,
$M_3=1.727$ GeV\cite{102}, and use the average value of 194 MeV
for the decay momentum $q_{2\eta\eta^\prime}$\cite{A5} since $f_2$ lies
very near to the threshold in the $\eta\eta^\prime$ decay mode\footnote{In this paper, the values of the
masses of other mesons are taken from Ref.\cite{PDG}}. In
fit a) the
parameters $x_j$, $y_j$
and $z_j$ are taken from Eq. (13) and in fit b)
$x_j$, $y_j$ are taken from Eqs. (14) and (15).
The parameters $\theta_1$, $\theta_2$,
$\theta_3$, $r_1$, $r_2$ and $r_3$ in two fits are determined as shown in
Table II and the predicted and the measured results are shown in Table
III. Comparing fit a) with fit b), we find that
three Euler angles and the predicted results
are not much altered, and that the $\chi^2$ of the
two fits are nearly equal, which shows that the new data on the
hadronic decays of $f_i$ into two pseudoscalar mesons are insensitive to
the possible glueball components of $\eta$ and $\eta^\prime$.

Based on the parameters with the lowest $\chi^2$,
the physical states $|f_1\rangle$, $|f_2\rangle$ and $|f_3\rangle$ can be
given by
\begin{eqnarray}
&&|f_1\rangle=-0.599|N\rangle+0.326|S\rangle-0.732|G\rangle,
\nonumber\\
&&|f_2\rangle=0.795|N\rangle+0.350|S\rangle-0.495|G\rangle,\\
&&|f_3\rangle=0.095|N\rangle-0.878|S\rangle-0.469|G\rangle.
\nonumber
\end{eqnarray}

From Eq. (16), one also can obtain
\begin{eqnarray}
&\Gamma(f_1\rightarrow\gamma\gamma):\Gamma(f_2\rightarrow\gamma\gamma):
\Gamma(f_3\rightarrow\gamma\gamma)=\nonumber\\
&M^3_1(5x_1+\sqrt{2}y_1)^2:M^3_2(5x_2+\sqrt{2}y_2)^2:M^3_3(5x_3+\sqrt{2}y_3)^2=
14.50:67.75:3.02.
\end{eqnarray}
This prediction can provide a test for the consistency of our results.

\section{Discussions}
\indent

Now we wish to discuss the properties of the mass matrix which can be used
to describe the mixing of the scalar mesons based on our preferred
results. In the $|N\rangle$, $|S\rangle$ and $|G\rangle$ basis, the
general form of
the mass matrix $M$ describing the mixing of the quarkonia and a glueball can be
written as\cite{matrix}
\begin{equation}
M=\left(\begin{array}{ccc}
M_N+2A_1&\sqrt{2}A_2&\sqrt{2}B_1\\
\sqrt{2}A_2&M_S+A_3&B_2\\
\sqrt{2}B_1&B_2&M_G
\end{array}\right),
\end{equation}
where $M_N$,
$M_S$ and $M_G$ represent the masses of
the bare states $|N\rangle$, $|S\rangle$ and $|G\rangle$, respectively;
$A_1$ ( $A_3$ ) is the amplitude of $|N\rangle$ (
$|S\rangle$ ) annihilation and reconstruction via intermediate gluons
states; $A_2$ is the amplitude of the transition between $|N\rangle$
and $|S\rangle$; $B_1$ ( $B_2$ ) is the amplitude of the transition between
$|N\rangle$ ( $|S\rangle$ ) and $|G\rangle$. If $A_1$, $A_2$ and
$A_3$ are set to
be zero, and $B_1$ is assumed to be equal to $B_2$, Eq. (16) would reduced
to the form employed in Ref.\cite{Wein}.

The physical states $|f_1\rangle$, $|f_2\rangle$ and $|f_3\rangle$ are
assumed to be the eigenvectors of the mass matrix $M$ with the
eigenvalues of $M_1$, $M_2$ and $M_3$, then we can have
\begin{equation}
UMU^\dagger=\left(\begin{array}{ccc}
M_1&0&0\\
0&M_2&0\\
0&0&M_3
\end{array}\right),~~
\left(\begin{array}{c}
|f_1\rangle\\
|f_2\rangle\\
|f_3\rangle
\end{array}\right)=
U\left(\begin{array}{c}
|N\rangle\\
|S\rangle\\
|G\rangle
\end{array}\right).
\end{equation}
Comparing Eq. (16) with Eq. (19), we have
\begin{equation}
U=\left(\begin{array}{ccc}
-0.599&0.326&-0.732\\
0.795&0.350&-0.495\\
0.095&-0.878&-0.469
\end{array}\right).
\end{equation}
Then the numerical form of the mass matrix can be given by
\begin{equation}
M=U^\dagger\left(\begin{array}{ccc}
M_1&0&0\\
0&M_2&0\\
0&0&M_3
\end{array}\right)U=\left(\begin{array}{ccc}
~1.436&0.018&-0.093\\
~0.018&1.656&~0.138\\
-0.093&0.138&~1.450
\end{array}\right).
\end{equation}

Eq. (21) shows that $A_2$ is very small. If $A_1$ and $A_3$ also can be
expected to be
very small, the mass level order of the bare states $|N\rangle$, $|S\rangle$ and $|G\rangle$
would be $M_S>M_G>M_N$, which is consistent with the argument
given by Refs.\cite{4241,CLOSE} while disagrees with the prediction that
the glueball state has a higher mass than the $q\bar{q}$ state\cite{Lat}.
Otherwise, the mass level order of $M_N$, $M_S$ and $M_G$ in scalar sector
would remain unclear. In addition, Eq. (21) implies that the
mass of the pure scalar glueball is about $1.5$ GeV, which is consistent
with the lattice QCD prediction\cite{Lat1}.

A salient property of Eq. (21) is that $B_1<0$ and $B_2>0$.
This shows that the amplitude of the transition between $|N\rangle$ and
$|G\rangle$ is negative while
the amplitude of the transition between $|S\rangle$ and
$|G\rangle$ is positive, which disagrees
with the assumption that $B_1=B_2$ in the model\cite{Wein}. In fact, in
the scalar sector, $B_1$ and $B_2$ should be nonperturbative effects dominantly,
there are not convincing reasons to expect that
the relation between $B_1$ and $B_2$ should behave as $B_1=B_2$.

We note that the values of $r_1$ and $r_2$ are inconsistent with that $r_1$
and $r_2$ should be less than the unit,
the prediction given by the perturbative theory.
We find that if we restrict that $r_1$, $r_2$ and $r_3$ in
the viewpoint of the perturbative theory, i.e., $r_1<1$, $r_2<1$ and $r_3<1$,
the $\chi^2$ increases from 2.05 to 3.80, but the results given above are not
much altered. However, in the scalar sector, there are not
any convincing reasons to expect that the perturbative theory should be valid.
The values of $r_1$ and $r_2$ imply that the nonperturbative
effects in the scalar sector could be rather large.

\section{Summary and conclusions}
\indent

Using three Euler angles, we introduce a mixing scheme to describing the
mixing of the isoscalar scalar mesons. In this mixing framework,
considering the four coupling modes as shown in Fig. 1, we construct the
effective Hamiltonian to investigate the two-body hadronic decays of
$f_0(1370)$, $f_0(1500)$ and $f_0(1710)$. The glueball-quarkonia content
of the three states is determined from the fit to the new data about
the hadronic decays of the three states presented by the WA102
collaboration. Our conclusions are as follows:

1). The large mixing effect exist in the three states.

1). The new data about the hadronic decays of $f_0(1370)$, $f_0(1500)$ and
$f_0(1710)$
are insensitive to the possible glueball components of $\eta$ and
$\eta^\prime$.

3). The nonperturbative effects in the scalar sector are rather large.

4). Our preferred results don't support the assumption employed by Weingarten's mass matrix
describing the mixing of the isoscalar scalar states\cite{Wein} that $B_1=B_2$.

\section{Acknowledgments}
\indent

This project is supported by the National Natural
Science Foundation of China under Grant Nos. 19991487 and 19835060, and
the Foundation of Chinese Academy of Sciences under Grant No. LWTZ-1298.

\newpage
\section*{Appendix A: \\
Formulae for the electormagnetic decays widths rates involving $\eta$ and $\eta^\prime$}
\begin{eqnarray}
&&\frac{\Gamma(\eta\rightarrow\gamma\gamma)}
{\Gamma(\pi^0\rightarrow\gamma\gamma)}=\frac{1}{9}
\left(\frac{M_\eta}{M_{\pi^0}}\right)^3
(5x_\eta+\sqrt{2}y_\eta)^2,\\
&&\frac{\Gamma(\eta^\prime\rightarrow\gamma\gamma)}
{\Gamma(\pi^0\rightarrow\gamma\gamma)}=\frac{1}{9}
\left(\frac{M_{\eta^\prime}}{M_{\pi^0}}\right)^3
(5x_{\eta^\prime}+\sqrt{2}y_{\eta^\prime})^2,\\
&&\frac{\Gamma(\rho\rightarrow\eta\gamma)}
{\Gamma(\omega\rightarrow\pi^0\gamma)}=
\left[
\frac{(M^2_\rho-M^2_\eta)M_\omega}
{(M^2_\omega-M^2_{\pi^0})M_\rho}
\right]^3x^2_\eta,\\
&&\frac{\Gamma(\eta^\prime\rightarrow\rho\gamma)}
{\Gamma(\omega\rightarrow\pi^0\gamma)}=
3\left[
\frac{(M^2_{\eta^\prime}-M^2_\rho)M_\omega}
{(M^2_\omega-M^2_{\pi^0})M_{\eta^\prime}}
\right]^3x^2_{\eta^\prime},\\
&&\frac{\Gamma(\phi\rightarrow\eta\gamma)}
{\Gamma(\omega\rightarrow\pi^0\gamma)}
=\frac{4}{9}\frac{m^2_u}{m^2_s}\left[
\frac{(M^2_\phi-M^2_\eta)M_\omega}{(M^2_\omega-M^2_{\pi^0})M_\phi}
\right]^3 y^2_\eta,\\
&&\frac{\Gamma(\phi\rightarrow\eta^\prime\gamma)}
{\Gamma(\omega\rightarrow\pi^0\gamma)}=
\frac{4}{9}\frac{m^2_u}{m^2_s}
\left[\frac{(M^2_\phi-M^2_{\eta^\prime})M_\omega}
{(M^2_\omega-M^2_{\pi^0})M_\phi}\right]^3y^2_{\eta^\prime},\\
&&\frac{\Gamma(J/\psi\rightarrow\rho\eta)}
{\Gamma(J/\psi\rightarrow\omega\pi^0)}
=\left[\frac{\sqrt{[M^2_{J/\psi}-(M_\rho+M_\eta)^2][
M^2_{J/\psi}-(M_\rho-M_\eta)^2]}}
{\sqrt{[M^2_{J/\psi}-(M_\omega+M_{\pi^0})^2][M^2_{J/\psi}-(M_\omega-M_{\pi^0})^2]}}\right]^3x^2_\eta,\\
&&\frac{\Gamma(J/\psi\rightarrow\rho\eta^\prime)}
{\Gamma(J/\psi\rightarrow\omega\pi^0)}
=\left[\frac{\sqrt{[M^2_{J/\psi}-(M_\rho+M_{\eta^\prime})^2][M^2_{J/\psi}-(M_\rho-M_{\eta^\prime})^2]}}
{\sqrt{[M^2_{J/\psi}-(M_\omega+M_{\pi^0})^2][M^2_{J/\psi}-(M_\omega-M_{\pi^0})^2]}}\right]^3
x^2_{\eta^\prime},
\end{eqnarray}
where $M_\rho$, $M_\omega$, $M_\phi$ and $M_{J/\psi}$ are the masses of
$\rho$, $\omega$, $\phi$ and $J/\psi$, respectively; $m_u$ and
$m_s$ are the masses of the constituent quark $u$ and $d$, respectively. Here we
take $m_u/m_s=0.642$ used in Ref.\cite{rosner}.

\newpage
\begin{table}
\begin{center}
\begin{tabular}{ccccccc}\hline
 &              &Fit1         &Fit2     \\
 &Exp.\cite{PDG}&$z_j\not=0$~($j=\eta$, $\eta^\prime$)&$z_j=0$~($j=\eta$, $\eta^\prime$)\\
                   &              &$\chi^2=1.64$    &$\chi^2=9.19$ \\
\hline
$\frac{\Gamma(\eta\rightarrow\gamma\gamma)}
{\Gamma(\pi^0\rightarrow\gamma\gamma)}$&$58.46\pm9.03$&53.76&63.67\\
$\frac{\Gamma(\eta^{\prime}\rightarrow\gamma\gamma)}
{\Gamma(\pi^0\rightarrow\gamma\gamma)}$&$540.78\pm104.44$&561.33&728.20\\
$\frac{\Gamma(\rho\rightarrow\eta\gamma)}
{\Gamma(\omega\rightarrow\pi^0\gamma)}$&$0.051\pm0.023$&0.066&0.073\\
$\frac{\Gamma(\eta^\prime\rightarrow\rho\gamma)}
{\Gamma(\omega\rightarrow\pi^0\gamma)}$&$0.086\pm0.016$&0.086&0.111\\
$\frac{\Gamma(\phi\rightarrow\eta\gamma)}
{\Gamma(\omega\rightarrow\pi^0\gamma)}$&$0.078\pm0.010$&0.074&0.066\\
$\frac{\Gamma(\phi\rightarrow\eta^\prime\gamma)}
{\Gamma(\omega\rightarrow\pi^0\gamma)}$&$0.0007\pm0.0005$&0.0003&0.0004\\
$\frac{\Gamma(J/\psi\rightarrow\rho\eta)}
{\Gamma(J/\psi\rightarrow\omega\pi^0)}$&$0.460\pm0.120$&0.482&0.533\\
$\frac{\Gamma(J/\psi\rightarrow\rho\eta^\prime)}
{\Gamma(J/\psi\rightarrow\omega\pi^0)}$&$0.250\pm0.079$&0.223&0.285\\\hline
\end{tabular}
\vspace{0.5cm}
\caption{The predicted and measured results of electromagnetic decays involving $\eta$,
$\eta^\prime$.}
\end{center}
\end{table}
\vspace{1cm}

\begin{table}
\begin{center}
\begin{tabular}{cccccccc}\hline

&$\chi^2$&$r_1$&$r_2$&$r_3$&$\theta_1$&$\theta_2$&$\theta_3$\\\hline
Fit a)&2.05&1.0&3.4&0.33&$-146^\circ$&$118^\circ$&$-151^\circ$\\
Fit b)&2.15&1.0&0&0.7&$-148^\circ$&$115^\circ$&$-146^\circ$\\\hline
\end{tabular}
\vspace{0.5cm}
\caption{The parameters determined from the fit.}
\end{center}
\end{table}

\begin{table}
\begin{center}
\begin{tabular}{cccc}\hline
&Exp.\cite{102}  &Fit a)                       &Fit b)\\
&              &$\chi^2=2.05$&$\chi^2=2.15$\\
\hline
$\frac{\Gamma(f_0(1370)\rightarrow\pi\pi)}{\Gamma(f_0(1370)\rightarrow
K\bar{K})}$&
$2.17\pm0.90$&2.453&2.397\\
$\frac{\Gamma(f_0(1370)\rightarrow\eta\eta)}{\Gamma(f_0(1370)\rightarrow
K\bar{K})}$&
$0.35\pm0.30$&0.248&0.314\\
$\frac{\Gamma(f_0(1500)\rightarrow\pi\pi)}{\Gamma(f_0(1500)\rightarrow
\eta\eta)}$&
$5.56\pm0.93$&5.581&5.853\\
$\frac{\Gamma(f_0(1500)\rightarrow
K\bar{K})}{\Gamma(f_0(1500)\rightarrow\pi\pi)}$&
$0.33\pm0.07$&0.335&0.308\\
$\frac{\Gamma(f_0(1500)\rightarrow\eta\eta^\prime)}
{\Gamma(f_0(1500)\rightarrow\eta\eta)}$&
$0.53\pm0.23$&0.528&0.484\\
$\frac{\Gamma(f_0(1710)\rightarrow \pi\pi)}{\Gamma(f_0(1710)\rightarrow
K\bar{K})}$&
$0.20\pm0.03$&0.191&0.200\\
$\frac{\Gamma(f_0(1710)\rightarrow\eta\eta)}{\Gamma(f_0(1710)\rightarrow
K\bar{K})}$&
$0.48\pm0.19$&0.230&0.223\\
$\frac{\Gamma(f_0(1710)\rightarrow\eta\eta^\prime)}
{\Gamma(f_0(1710)\rightarrow K\bar{K})}$&
$<0.04$($90\%$CL)&0.035&0.021\\\hline
\end{tabular}
\vspace{0.5cm}
\caption{The predict and  measured results of the
hadronic decays of $f_i$.}
\end{center}
\end{table}

\begin{figure}
\epsfxsize=15cm \epsfbox{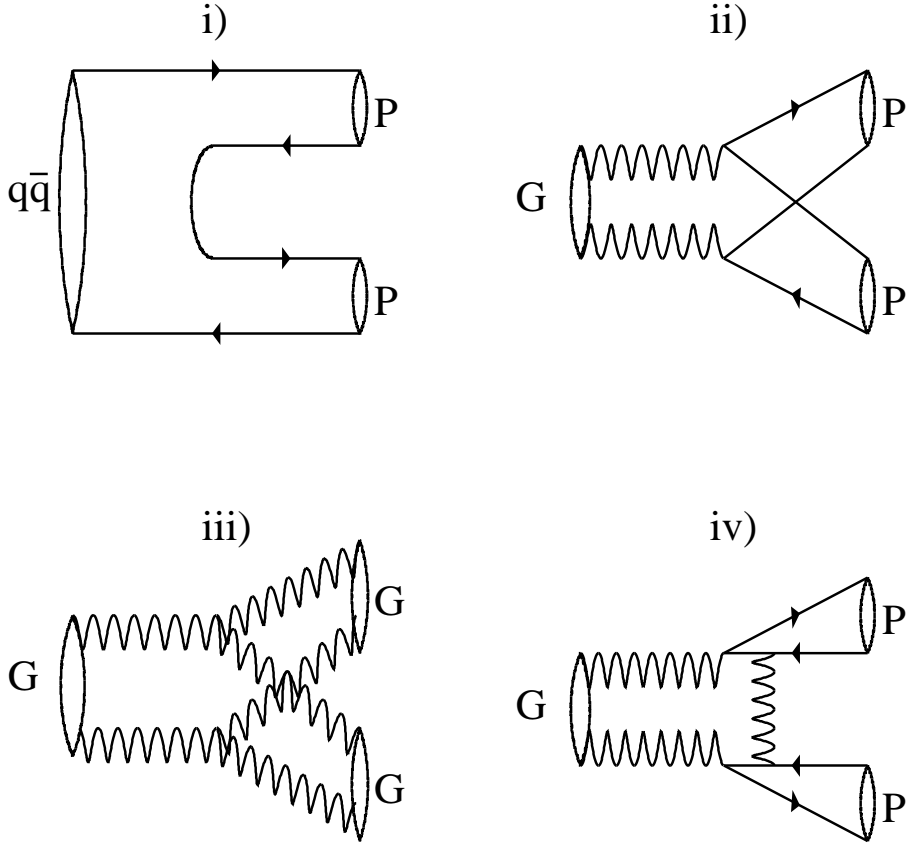}
\caption{The coupling modes considered in this work. i) The
coupling of the quarkonia components of $f_i$ to the final pseudoscalar
meson pairs, ii) the coupling of the glueball components of $f_i$ to the final pseudoscalar
meson pairs via $qq\bar{q}\bar{q}$ intermediate states, iii) the coupling
of the glueball components of $f_i$ to the glueball components of the final
isoscalar pseudoscalar mesons, and iv) the coupling of the glueball
components of $f_i$ to the quarkonia of the final isoscalar pseudoscalar meson pairs.}
\end{figure}

\end{document}